\documentclass[useAMS,usenatbib]{mn2e}
\usepackage{epsfig}

\begin{document}

\title{Dynamics of clumps embedded in a hot accretion flow with toroidal magnetic field}

\author[F. Khajenabi , M. Rahmani \& S. Abbassi]{Fazeleh Khajenabi $^{1}$\thanks{E-mail:
f.khajenabi@gu.ac.ir;} and Mina Rahmani$^{2}$ and Shahram Abbassi$^{3,4}$ \\
$^{1}$ Department of Physics, Faculty of Sciences, Golestan University, Gorgan 49138-15739, Iran\\
$^{2}$ Department of Physics, Faculty of Sciences, Damghan University, Damghan, Iran\\
$^{3}$ Department of Physics, Faculty of Sciences, Ferdowsi University, P.O. Box 1436, Mashhad, Iran\\
$^{4}$ School of Astronomy, Institute for Research in Fundamental Sciences (IPM), P.O. Box 19395-5531, Tehran, Iran}

\maketitle

\date{Received ______________ / Accepted _________________ }

\begin{abstract}
Dynamics of clumps within a magnetized advection dominated accretion flow is investigated by solving the collisionless  Boltzmann equation and considering the drag force due to the relative velocity between the clumps and the gas. Toroidal component of the magnetic field is assumed to be dominant. Dynamical properties of the hot gaseous component such as the radial and the rotational velocities are affected by the magnetic effects, and so, drag force varies  depending on the strength of magnetic field and the velocity dispersion of the clumps is then modified significantly. We show that when magnetic pressure is less than the gas pressure, the root of the averaged radial velocity square of the clumps decreases at the inner parts of the hot flow and increases slightly at its outer edge.

\end{abstract}


\section{Introduction}
While the standard    model of the accretion discs (Shakura \& Sunyaev 1973) and the alternatives models  (e.g., Narayan \& Yi 1994; Blandford \& Begelman 1999;  Narayan, Igumenshchev \& Abramowicz  2000) have been widely used  for describing  commonly supposed  one-component gaseous accreting systems, there are growing theoretical and observational evidences for the co-existence of clumps with a gaseous medium in a number of the accreting objects. The broad line emission in the spectrum of radio galaxies, quasars and some of the observational features of AGNs have been explained based on such two-phase models (e.g., Rees 1987; Kuncic, Blackman \& Rees 1996; Celotti \& Rees 1999; Malzac \& Celotti 2002; Barai, Proga \& Nagamine 2012). Structure of NGC 1068 as one of the best-studied Seyfert II galaxies has been studied based on a clumpy accretion model, in which accretion arises because of the gravitational interaction between the clumps  (Kumar 1999). These clumps not only may affect  dynamics of the gaseous accreting component, but they  can modify radiative appearance of the system. On the other hand, understanding  mechanism of the instabilities  which may lead to the formation of cold clouds or clumps within an accreting object is also challenging (e.g., Krolik 1998; Gammie 1998; Blaes \& Socrates 2003). Part of the difficulties  comes from the lack of a clear understanding  about physics of a purely gaseous disc, in particular mechanisms of angular momentum transport, energy transport and possible roles of the magnetic field or self-gravity. In theories for a two-phase accreting system, there are  challenging questions not only about nature of the accretion  process itself, but it is desirable to understand  internal structure of the clumps and their interactions with the intercloud  gaseous medium.

Although accretion onto a black hole is likely to be very chaotic, in a two-phase model for realizing such systems a common assumption is that the clumps are in pressure equilibrium with their surrounding gaseous medium. Sustaining temperature and density gradients  to maintain a multi-phase medium is explainable to some extend by the combined action  of the  magnetic effects and the radiation field. In fact, magnetic field can play an important role in confinement of clumps close to the central region of AGNs (Rees 1987). According to a study by Celotti \& Rees (1999), co-existence of clumps with a magnetized hot flow may affect observed spectral signatures of advection dominated accretion flows (ADAFs; Narayan \& Yi 1994). In ADAFs, generated heat by the turbulence is advected inwards with the flow rather being radiated out of the system.

Despite of the important  dynamical role of the magnetic effects, none of the previous works have studied dynamics of the clumps in a hot {\it magnetized} accretion flow. However, Wang, Cheng \& Li (2012) (hereafter WCL) studied  dynamics of the clumps embedded in the non-magnetized ADAFs by solving the collisionless Boltzmann equation analytically. They showed that the velocity dispersion of clumps is one magnitude higher than the ADAF when the coupling between the clumps and the intercloud medium is strong.

In this work, we extend model of WCL to a magnetized case, in which clumps are embedded in a magnetized hot accretion flow. For simplicity, it is assumed that the toroidal component of the magnetic field is dominant. In the next section, Boltzman equation for describing dynamics of the slumps and the general solutions for the hot accretion flow are presented. In section 3, we explore role of the various input parameters in the dynamics of clumps. The final section is devoted to concluding points.

\section{General formulation}
For describing dynamics of the clumps in the accreting gas, we follow WCL approach where the collisionless Boltzman equation is solved analytically subject to some simplifying assumptions. We introduce the distribution function $F$, where ${\cal F}=\Delta N/R\Delta R\Delta z\Delta \phi \Delta v$.  We can write Boltzman equation in the cylindrical coordinate $(R, \phi, z)$, i.e.
\begin{displaymath}
\frac{\partial {\cal F}}{\partial t}+\dot R\frac{\partial {\cal F}}{\partial R}+\dot \phi\frac{\partial {\cal F}}{\partial \phi}+\dot z \frac{\partial F}{\partial z}+\dot v_{R} \frac{\partial {\cal F}}{\partial v_{R}}+\dot v_{\phi}\frac{\partial {\cal F}}{\partial v_{\phi}}
\end{displaymath}
\begin{equation}
 +\dot v_{z} \frac{\partial {\cal F}}{\partial v_{z}}+{\cal F} (\frac{\partial \dot v_{R}}{\partial v_{R}}+\frac{\partial \dot v_{\phi}}{\partial v_{\phi}}+\frac{\partial \dot v_{z}}{\partial v_{z}} )=0.
\end{equation}
where $\dot R=v_{R}$, $\dot \Phi={ v_{\phi}}/{R}$ and $\dot z=v_{z}$. We assume the system is axisymmetric. Also, the drag force which depends on the relative velocity between the clumps and the gaseous medium has two components. Thus, we have
\begin{displaymath}
\frac{\partial {\cal F}}{\partial t}+v_{R}\frac{\partial {\cal F}}{\partial R}+v_{z}\frac{\partial {\cal F}}{\partial z}+(\frac{v_{\phi}^2}{R}-\frac{\partial \Phi}{\partial R}+F_{R})\frac{\partial {\cal F}}{\partial v_{R}}
\end{displaymath}
\begin{displaymath}
 +(F_{\phi}-\frac{v_{R}v_{\phi}}{R})\frac{\partial {\cal F}}{\partial v_{\phi}}-\frac{\partial \Phi}{\partial z}\frac{\partial {\cal F}}{\partial v_{z}}+2 {\cal F}[F_{\phi}(v_{\phi}-V_{\phi})
\end{displaymath}
\begin{equation}
+F_{R}(v_{R}-V_{R})]=0.
\end{equation}
where the gravitational potential of the central object is $\Phi=GM/(R^2+z^2)^{1/2 }$, and the components of the drag force are $F_{R}=f_{R}(v_{R}-V_{R})^2$ and $F_{\phi}=f_{\phi}(v_{\phi}-V_{\phi})^2$. Here,  $V_{R}$ and $V_{\phi}$ are the radial and the rotational velocities of the ADAF. The drag forces in the $z-$direction has been neglected. In this study,  we assume that there is strong coupling between clumps and the gaseous medium and so we can write $\langle v_{R}\rangle =V_{R}$ and $\langle v_{\phi} \rangle =R \Omega_{A}$ where $\Omega_{A}$ is the angular  velocity of the ADAF.

Upon substituting dynamical profiles of the ADAF into the Boltzman equation, the root mean radial velocity square of the clumps $\langle v_{\rm R}^2 \rangle ^{1/2}$ is obtained analytically (WCL):
\begin{displaymath}
\langle v_{\rm R}^2 \rangle =c^2 [\frac{1}{2} [\alpha^2 c_{1}^2 \Gamma_{\rm R} \Lambda_{\frac{3}{2}} + (1-c_{2}^2) \Lambda_{\frac{5}{2}} - \frac{\alpha c_1 c_2 }{2\Gamma_{\phi} \Lambda_{\frac{7}{2}}}]
\end{displaymath}
\begin{equation}
+ \frac{V_{\rm out}^2}{c^2 r_{\rm out}^{1/2}} \exp (-\Gamma_{\rm R} r_{\rm out}) ] r^{1/2} \exp (\Gamma_{\rm R} r),
\end{equation}
where $c_1$ and $c_2$ are coefficients of the radial and the rotational velocities of the ADAF. At the outer edge $R_{\rm out}$, it was assumed $\langle v_{\rm R}^2 \rangle =V_{\rm out}^2$. Moreover, the coefficients of the drag force are replaced by $\Gamma_{\rm R} = f_{\rm R} R_{\rm Sch}$ and $\Gamma_{\phi} = f_{\phi} R_{\rm Sch}$. Function $\Lambda_{q}$ has been introduced by WCL as $\Lambda_{q}=\int_{r}^{r_{\rm out}} x^{-q} \exp (-\Gamma_{\rm R} x) dx$.

In WCL,  coefficients $c_1$ and $c_2$ have been substituted based on the standard non-magnetized ADAF solution (Narayan \& Yi 1994). Here, we substitute these coefficients from the magnetized ADAF solution by  Akizuki \& Fukue (2006). They assumed that the dominant component of the magnetic field is toroidal and the viscosity is proportional to the sum of the magnetic pressure and the gas pressure. The constant of proportionality is $\alpha$. Magnetized self-similar solutions are written as
\begin{equation}
V_{\rm R} = - c_1 \alpha \sqrt{\frac{GM}{R}},
\end{equation}
\begin{equation}
V_{\rm \phi} = c_2 \sqrt{\frac{GM}{R}},
\end{equation}
\begin{equation}
C_{\rm s}^2 = c_3 \frac{GM}{R},
\end{equation}
where $C_{\rm s}$ is the sound speed. The coefficients $c_{1}$, $c_{2}$ and $c_{3}$ are obtained (Akizuki \& Fukue 2006):
\begin{displaymath}
 c_{1}=\frac{1}{3 \alpha^2 (1+\beta)}[\sqrt{ (\frac{1-s}{1+s}-\beta+ 3 \epsilon )^2+18 \alpha^2(1+\beta)^2}
\end{displaymath}
\begin{equation}
-(\frac{1-s}{1+s}-\beta+ 3 \epsilon )],
\end{equation}
\begin{displaymath}
 c_{2}^2=\frac{\epsilon}{3 \alpha^2 (1+\beta)^2 }[\sqrt{ (\frac{1-s}{1+s}-\beta+ 3 \epsilon )^2+18 \alpha^2 (1+\beta)^2}
\end{displaymath}
\begin{equation}
-(\frac{1-s}{1+s}-\beta+ 3 \epsilon )],
\end{equation}
\begin{displaymath}
 c_{3}=\frac{1}{9 (1+s) \alpha^2(1+\beta)^2}\times
\end{displaymath}
\begin{equation}
[\sqrt{ (\frac{1-s}{1+s}-\beta+ 3 \epsilon )^2+18 \alpha^2 (1+\beta)^2}-(\frac{1-s}{1+s}-\beta+ 3 \epsilon )],
\end{equation}
where
\begin{equation}
\epsilon=\frac{23-\gamma}{9 \gamma-1}\frac{1}{f},
\end{equation}
and $\beta$ is the ratio of the magnetic pressure to the gas pressure, i.e. $\beta=\frac{p_{mag}}{p_{gas}}$. Also, $\gamma$ is the adiabatic index and $f$ and $s$ are  the advection parameter and the mass loss parameter, respectively.

\section{Analysis}
Having analytical solutions for the gas component and an analytical relation for the root mean radial velocity square of the clumps, namely $\langle v_{\rm R}^2 \rangle ^{1/2}$, as a function of the radial distance, we can do a parameter analysis to clarify possible magnetic effects.  In all subsequent Figures,  mass of the central object is assumed to be one solar mass, coefficient of viscosity is $\alpha =0.1$ and the adiabatic index is $\gamma = 1.4$. The coefficients of the drag force are fixed   as $\Gamma_{\rm R}=5\times 10^{-2}$ and $\Gamma_{\varphi} = 2.8 \times 10^{-3}$.  Obviously, the non-magnetic case corresponds to $\beta=0$ and as this ratio increases, the system tends to be more magnetically dominated.

\begin{figure}
\epsfig{figure=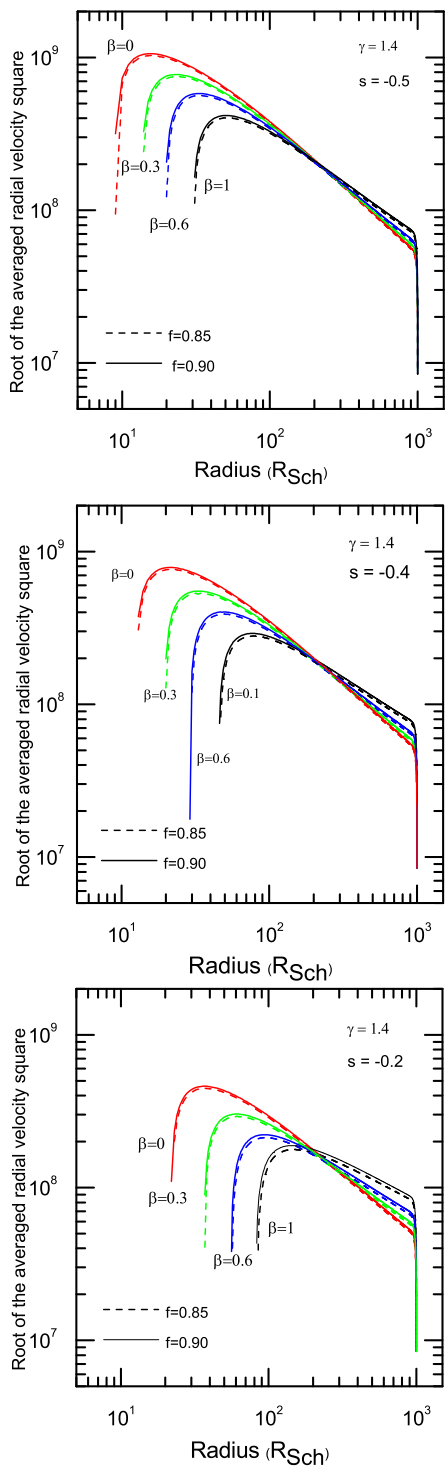,angle=0,scale=1.0}
\caption{Root of the averaged radial velocity square $\langle v_{R}^2 \rangle ^{1/2}$ of clumps versus the radial distance for a central object with one solar mass. Values of $s$ are $-0.5$, $-0.4$ and $-0.2$ and also we have $\gamma =1.4$ and $\alpha=0.1$. Different values for the ratio of the magnetic to the thermal pressure $\beta$ are considered and the solid and the dashed curves are corresponding to $f=0.9$ and $0.85$, respectively.}
\label{fig:f1}
\end{figure}

\begin{figure}
\epsfig{figure=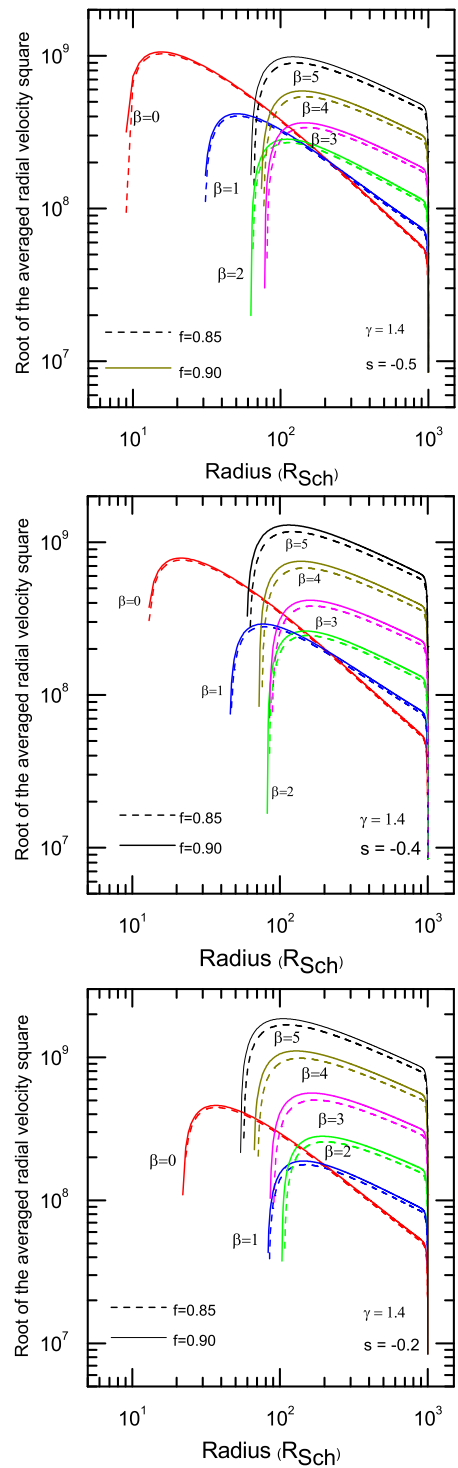,angle=0,scale=1.0}
\caption{Same as Figure 1, but  magnetically dominated cases with $\beta$ larger than one are also shown.}
\label{fig:f2}
\end{figure}

\begin{figure}
\epsfig{figure=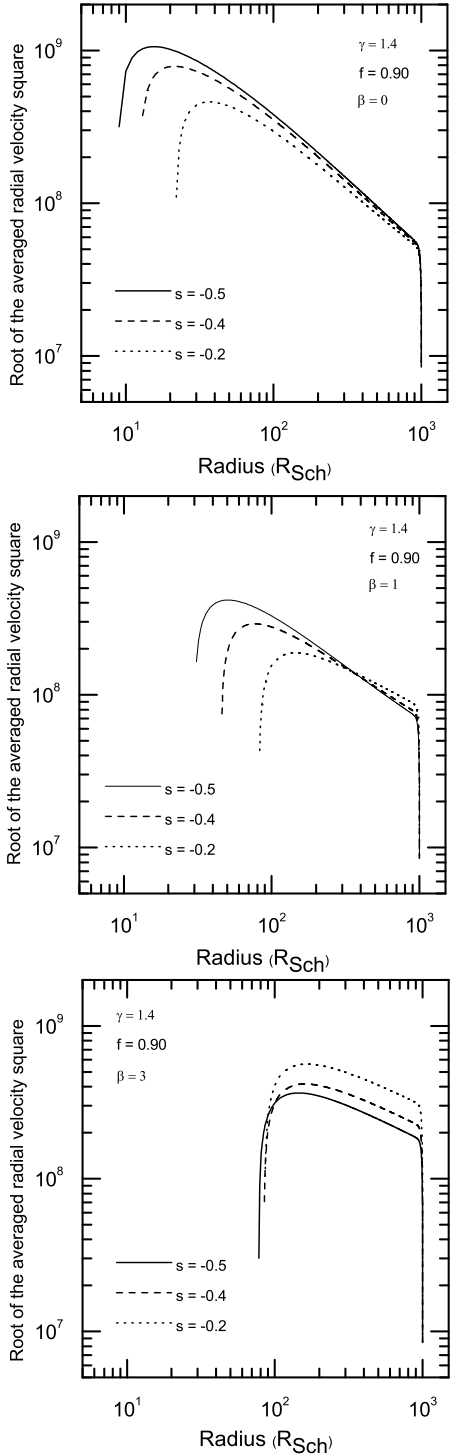,angle=0,scale=1.0}
\caption{Effects of the parameter $s$ are explored for different values of $\beta$. Mass of the central object is one solar mass and we have $\gamma = 1.4$ and $f=0.9$.}
\label{fig:f3}
\end{figure}

\begin{figure}
\epsfig{figure=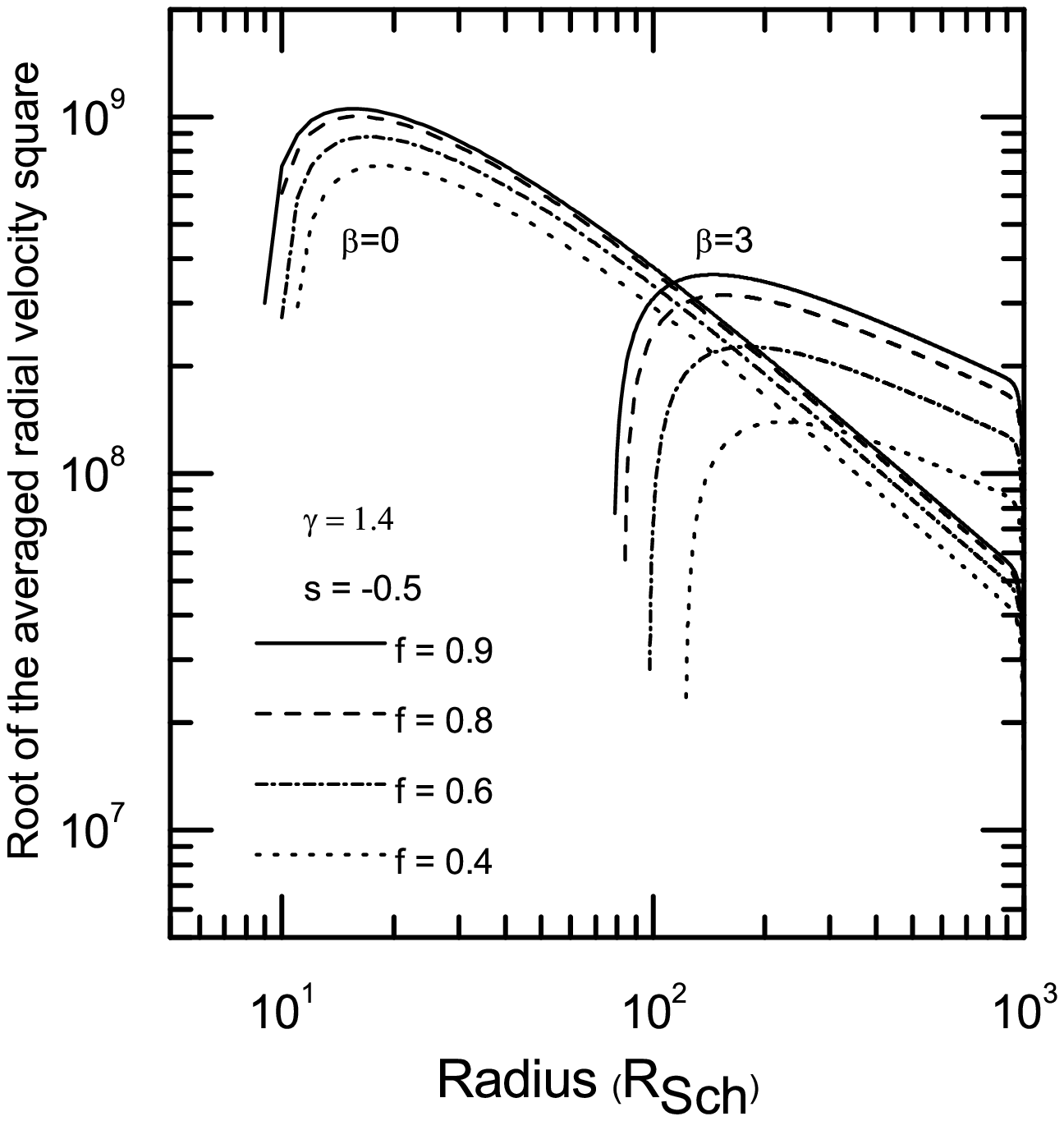,angle=0,scale=0.5}
\caption{Effects of the advected energy fraction $f$ are examined for the non-magnetized and magnetically dominated cases. Again, mass of the central object is one solar mass and we have $\gamma = 1.4$, $s=-0.5$ and $\alpha=0.1$.}
\label{fig:f4}
\end{figure}

Figure 1 shows profile of the root mean radial velocity square of the clumps as a function of the radial distance normalized by $R_{\rm Sch}$. Solid and dashed curves are corresponding to $f=0.9$ and $0.85$, respectively. As long as the gas component is fully advective, the results are not sensitive to the variations of advection parameter $f$, irrespective of the magnetic effects. No-wind solution corresponds to $s=-0.5$. But this case and  other cases with $s=-0.4$ and $s=-0.2$ are strongly affected because of considering magnetic effects. The value of $\langle v_{\rm R}^2 \rangle ^{1/2}$ reduces in the inner parts of the disc, but it increases at the outer parts of the system. This trend becomes more significant as the ratio $\beta$ increases. In other words, the slope of the curves becomes shallower as the system becomes more magnetically dominated. Strongly magnetized cases with $\beta$ larger than one are also investigated in Figure 2. In these cases, behavior of the results are similar to the less magnetically situations, though it is unlikely to have ADAFs with such a strong magnetic field. But in magnetically dominated cases, the effect of advected energy fraction $f$ becomes more significant in comparison to, say, non-magnetized case. In such cases, as the parameter $f$ reduces which means more energy is radiated, the root mean radial velocity square of the clumps reduces as well.

Effect of variations of the parameter $s$ is shown in Figure 3. The ratio $\beta$ is assumed to be $0$ (top), $1$ (middle) and $3$ (bottom). In the non-magnetized case, i.e. $\beta =0$, as the parameter $s$ increases the value of $\langle v_{\rm R}^2 \rangle ^{1/2}$ reduces. For no-wind case ($s=-0.5$), the root mean radial velocity square  is larger than the cases with outflows. Reduction to the value of $\langle v_{\rm R}^2 \rangle ^{1/2}$ because of the emergence of outflows is justified if we note that outflows extract mass, angular momentum and energy from the system. However, in our simplified model, just mass transfer by the outflows is taken into account. But as magnetic effects become stronger, we see a less reduction to the root mean radial velocity square of the clumps so that for a highly magnetic case like $\beta =3$, the value  of $\langle v_{\rm R}^2 \rangle ^{1/2}$ increases with the emergence of the outflows. In such a magnetically dominated case, dynamics of the gas flow is under control of the magnetic effects. Thus, the drag force is modified significantly. This in turn leads to an enhancement of the root mean radial velocity square of the clumps.

Finally, we examine effects of the advected energy fraction $f$ in Figure 4. Here, a non-magnetized case and a magnetically dominated case with $\beta =3$ are considered for different values of $f$. Whereas non-magnetic results are not very sensitive to the parameter $f$, in the magnetized case there is a significant reduction to the root mean radial velocity square of the clumps because of the magnetic effects.

Considering the above findings on the role of magnetic field, one may raise a question about the fate of the magnetized clumps. We note that our clumps are in pressure equilibrium with a hot plasma. Generally, not only the clumps but also the gas itself may couple to the field lines whilst maintaining total pressure, i.e. $p_{total} =p_{thermal}+p_{mag}$, continuity across the boundary of each clump. As it has already been discussed by Rees (1987) and Kuncic et al. (1996) among the others, magnetic field would support a clump against rapid expansion and dispersion. This result is independent of magnetic field geometry of the gas and the clumps where residing within them (for details see Kuncic et al. (1996)). Obviously, our analysis is not addressing the stability of magnetized clumps, but the main goal is to study dynamics of clumps within the gas considering the fact that magnetic fields would lead to more stable clumps. Note that the parameter $\beta$ in our model represents level of the magnetization of the gas not the clumps. In general, we have to describe magnetic fields within a clump by a different parameter $\beta$. Uncertainties in the complex physics involved in the magnetic field geometry within a clump forced us to consider the simplest case where the internal pressure is entirely thermal. However, the gas component is assumed to be magnetized.

Coupling between the clumps and the gas is provided by the drag force where the size of the clumps and the density of the gas and the relative velocity of clumps and the gas are  three main factors for determining the magnitude of this force. In our formulation, the relative velocity of the clumps and the gas appears explicitly. But dependence of the drag force to the size of a clump and the density of the gas is described by the parameters $f_{R}$ and $f_{\phi}$. In the non-magnetized case, these parameters are assumed to be constant by WCL and this assumption has been justified theoretically. We can also follow a similar approach but for a magnetized gas. As each clump moves inward it may undergoes contraction. Thus, we can expect to have hotter clumps because of the contraction. On the other hand, there is efficient cooling by the radiation which prevents each clump becomes too hotter than its ambient medium. Then, like WCL we assume temperature of the clumps has a weak dependence on the radial distance to the central object, i.e. $T_{cl} \propto R^{- \zeta}$ where $\zeta \ll 1$. Our clumps are in pressure equilibrium across their boundaries with the ADAF. In the magnetized case, the internal thermal pressure of each clump is equal to the total pressure of the medium, $p_{total}=p_{thermal}+p_{mag}$. Since the ratio of the magnetic pressure to the thermal pressure $\beta$ is constant in our model, we can write $p_{total} = (1+\beta )p_{thermal}$. Our aim is to analyze the radial dependence of the coefficients $f_{R,\phi}$. In our formulation, radial dependence of these coefficients in the magnetized case would be the same as the non-magnetized system because the parameter $\beta$ is constant. Pressure equilibrium of a clump with its surrounding medium implies its radius $R_{cl}$ scale with the radial distance as $R_{cl} \propto R^{(2-s-\zeta )/3}$. Assuming the mass of a clump does not change during its motion within the gas, radial dependence of its density becomes $n_{cl} \propto R^{\zeta +s-2}$. Density of the gas has a self-similar profile as $n_{gas} \propto R^{s-1}$. Since $f_{R,\phi} = n_{gas}/(n_{cl} R_{cl})$, we obtain $f_{R,\phi} \propto R^{(1+s-2\zeta )/3}$. WCL assumed $\zeta = 0.2$ and for $s=-0.5$ and $s=-0.2$, we obtain $f_{R,\phi}\propto R^{1/30}$ and $f_{R,\phi} \propto R^{4/30}$ respectively. Thus, radial dependence of the coefficients are weak and it is reasonable to take these parameters as constant.

In  Figures 1-4, we found the root mean radial velocity square  $\langle v_{R}^2 \rangle$ increases as the ratio $\beta$ increases. In the non-magnetized case, WCL showed $\delta = \langle v_{R}^2 \rangle / \langle v_{R}  \rangle \sim 10  $ and  since the capture rate of the clumps at the inner edge of the disc is directly proportional to this ratio (see eq. (34) of WCL), the capture rate would be very high. According to our Figures, the ratio $\delta$ increases in particular for $\beta >1$. When magnetic fields are neglected, clumps are dynamically less stable and various types of the instabilities like Kelvin-Helmholtz instability may lead to the disruption of a clump. Thus, some of the clumps before arriving to the inner edge of the accreting gas would not survive against strong dynamical forces unless magnetic confinement is considered. But our analysis shows that magnetic fields not only lead to a larger capture rate of the clumps at the inner edge, but also the clumps are dynamically more stable.  Note that self-similar solutions for the gas component is not valid at the boundaries.

Observations and theoretical considerations are supporting existence of winds or outflows in a hot accretion flow. Outflows generally not only extract mass and energy from the system but also it may carry angular momentum. These complexities prevented us to study inflow-outflow systems analytically. But in our simple model for the hot gaseous component, only mass loss by the outflows is considered via a single parameter $s$. Obviously, we are just considering the structure of the accreting flow even in the presence of outflows. Not only launching mechanisms are disregarded but also  structure of the outflows is beyond the scope of this simple model. For this reason, within the framework of our model, we can only study dynamics of the clumps within the accreting flow not the structure of a clumpy outflow. For $s=-0.5$ the mass of the accreting flow is conserved and there is no wind. But for other values of $s$, part of the mass of the accreting flow is lost by winds or outflows. However, just recently clumpy outflows from hot flows have been studied by doing numerical simulation (e.g., Takeuchi, Ohsuga \& Mineshige 2013; Ohsuga \& Mineshige 2011). Takeuchi, Ohsuga \& Mineshige (2013) studied properties of the clumpy outflows, in particular mechanisms of creating clumpy outflows. In their simulations, the outward radiation pressure overcomes inward gravity force and so, the system is subject to the Rayleigh-Taylor (RT) instability (also see, Jacquet \& Krumholz  2011). In other words, it seems mechanism of clumps formation is a radiation-induced RT instability. However, it is unlikely that RT instability is responsible for the formation of clumps within the accreting flow itself. Thermal instability is a plausible physical mechanism of creating clumpy accreting flow. However, magnetic field has a stabilizing role in a preferred direction when the system is thermally unstable. In fact, compression due to the existence of the magnetic field occurs only perpendicular to the magnetic line of force. So, one may expect more elongated clumps aligned to the magnetic fields within a magnetized accreting flow. On the other hand, heat is conducted mainly along the magnetic field lines. Although the adopted geometry for the clumps in our model is spherical, it seems that when magnetic field is considered the clumps would be elongated following the magnetic field geometry. Future numerical simulations can clarify these physical considerations.

\section{Conclusion}
We studied  dynamics of the clumps in a {\it magnetized} hot accretion flow. Our analysis is a direct generalization of WCL to the magnetic case, in which the toroidal component of the magnetic field is assumed to be dominant. We note that magnetic fields not only affect dynamics of the gas component, but also the shape and the confinement of clumps are also influenced by the magnetic effects, though these are not considered in our simplified model. Like WCL, drag force due to the relative velocity between the clumps and the gas flow is taken into account. Thereby, dynamics of clumps indirectly are affected by the magnetic effects.

A detailed parameter survey has been done and it was shown that the root mean radial velocity square of the clumps is significantly modified when magnetic effects on the gas component is considered. The ratio of the magnetic to the thermal pressure $\beta$ plays a vital role. When this ratio is less than one, the value of the root mean radial velocity square reduces at the inner parts of the disc, but this quantity increases at large radii. This trend is independent of the other input parameters. For magnetically dominated cases with $\beta$ larger than one, however, we found significant enhancement to the root mean radial velocity square of the clumps. Similarity solutions for the magnetized gas flow show that both the radial and the azimuthal components of the gas velocity are affected by the magnetic forces. It leads to some changes in the relative velocity between the clumps and the gas and the drag force. It would be interesting to generalize our model by including vertical and the radial  components of the magnetic field.


\section*{Acknowledgments}
We are grateful to the anonymous referee whose detailed and careful comments helped to improve the quality of this paper.


\vspace{0.75cm}

{\bf REFERENCES}

\noindent Barai P., Proga  D., Nagamine  K., 2012, MNRAS, 424, 728

\noindent Blandford  R. D., Begelman  M. C., 1999, MNRAS, 303, 1

\noindent Blaes O., Socrates A., 2003, ApJ, 596, 509

\noindent Akizuki C., Fukue J., 2006, PASJ, 58, 469

\noindent Begelman M. C.,

\noindent Celotti A., Fabian A. C., Rees M. J., 1992, MNRAS, 255, 419

\noindent Celotti A., Rees M. J., 1999, MNRAS, 305, L41

\noindent Gammie C., 1998, MNRAS, 297, 929

\noindent Jacquet E., Krumholz M. R., 2011, ApJ, 730, 116

\noindent Kumar P., 1999, ApJ, 519, 599

\noindent Kuncic Z., Blackman E. G., Rees M. J., 1996, MNRAS, 283, 1322

\noindent Krolik J., 1998, ApJ, 498, L13

\noindent Malzac J., Celotti A., 2002, MNRAS, 335, 23

\noindent Narayan R., Yi I., 1994, ApJ, 428, L13

\noindent Narayan R., Igumenshchev I. V., Abramowicz M. A., 2000, ApJ, 539, 798

\noindent Ohsuga K., Mineshige S., 2011, ApJ, 736, 2

\noindent Rees M. J., 1987, MNRAS, 228, 47

\noindent Shakura N. I., Sunyaev R. A., 1973, A\& A, 24, 337

\noindent Stella L., Rosner R., 1984, ApJ, 277, 312

\noindent Takeuchi S., Ohsuga K., Mineshige S., 2013, PASJ, 65, 88

\noindent Wang J.-M., Cheng C., Li Y.-R., 2012, ApJ, 748, 147 (WCL)

%

\bibliographystyle{mn2e}


%
%
%
\end{document}